\documentstyle[eqsecnum,aps]{revtex}
\input{psfig}
\begin{document}
\draft
\title{ Multipole analysis of spin observables in vector meson photoproduction}
\author{\c{C}etin \c{S}avkl\i \ and Frank Tabakin \cite{byline} }
\address{ Department of Physics \& Astronomy, University of Pittsburgh,
 Pittsburgh, Pennsylvania  15260}
\author{Shin Nan Yang }
\address{Department of Physics, National Taiwan University, Taipei, Taiwan
10764, ROC\cite{byline2} and\\
Nuclear Science Division MS70A-3307, Lawrence Berkeley National
Laboratory,
University of California, Berkeley, CA 94720.}
\date{\today}
\maketitle
\begin{abstract}
A multipole analysis of vector meson photoproduction is formulated as a
generalization of the pseudoscalar meson case. Expansion of spin observables
in the multipole basis and behavior of these observables near threshold
and resonances are examined.

\end{abstract}

\widetext
\section{Introduction}
\label{sec:introduction}
With the advent of CEBAF there is renewed interest in measuring the
photoproduction of vector mesons $(\rho,\omega,\phi)$ \cite{laget,snyang}. In
anticipation of such experiments, the general nature of spin observables for
vector mesons has been studied\cite{PichowskySavkliTabakin} using a helicity
amplitude approach.
  In this paper, that study is extended by introducing electric and magnetic
multipole amplitudes for the photoproduction of vector mesons. The multipoles
include the final state orbital angular momentum and thus, near threshold,
 provide a
natural truncation to relatively few multipoles. Also, since resonances have
definite $\ell$-values, the isolation of isobar dynamics occurs most
naturally
in a resonant multipole amplitude. Expressions for the full array of spin
observables in terms of these multipoles are derived. Then, general rules for
the angular dependence of these observables and their dependence on
 multipoles
are discussed. We hope that such rules will be helpful in analyzing
future experiments.

\section{Multipoles For vector meson production}
\begin{figure}
\centering{\psfig{figure=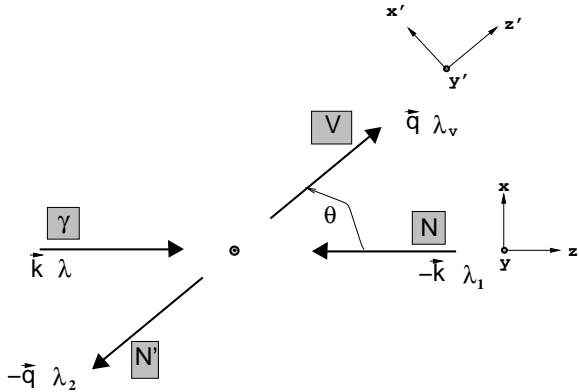,width=3.0in}}
\caption{The coordinate system and kinematical variables for vector meson
photoproduction. Here V denotes the vector meson
and $\lambda_V$ its helicity.}
\label{kinemat}
\end{figure}

The general structure of the scattering amplitude for the photoproduction of
vector mesons is:

\begin{eqnarray}
< \vec{q}\, \lambda_{V}\,m_2 \mid T \mid \vec{k}\,&\lambda_{\gamma}&
\, m_1 >\nonumber\\
&\equiv&\hat{\varepsilon}_{\lambda_{V}}^{*}(\vec{q}\,)\cdot <m_2 \mid
\stackrel{\leftrightarrow}{J} \mid  m_1 >\cdot
\hat{\varepsilon}_{\lambda_{\gamma}}(\vec{k}\,)\nonumber\\
&\equiv&<m_2 \mid {\cal J}_{\lambda_{V}\lambda_{\gamma}} \mid m_1>,
\label{tmatrix}
\end{eqnarray}
where $\vec{k},\lambda_{\gamma}$ and $\vec{q},\lambda_V$ denote the
three-momentum and helicity of the initial photon and final vector meson;
$m_1$ and $m_2$ are the initial and final nucleon spins quantized along the
$z$-axis, respectively ( see Fig.~\ref{kinemat}).  Isospin factors are
suppressed.
 The current tensor $\stackrel{\leftrightarrow}{J}$ is
defined in the Pauli-spinor space of the initial and final baryon. One can
express this current in terms of 12 generalized CGLN amplitudes\cite{cgln},
and proceed to find an expansion of those amplitudes in terms of
multipoles.\footnote{This course of action was taken in the analysis of
Ref.\cite{TabakinFasanoSaghai} for pseudoscalar mesons.} Reference \cite{pmc}
gives the transformation matrices connecting helicity amplitudes to a set
of gauge invariant covariant amplitudes for this reaction. In our paper, we
follow a direct approach to relate helicity amplitudes to multipoles, without
using CGLN amplitudes as an intermediate step.
\subsection{Multipole amplitudes}
The quantum numbers introduced in the multipole analysis are: the total
angular momentum $J$ and its projection along the $z$-axis $M,$ the relative
orbital angular momentum $\ell_{V}$ of the final particles;\footnote{We denote
the vector particle as $V,$ which refers to $\rho,$ $\omega$
or $\phi$ mesons. Here, $\ell_V$ is the vector meson-nucleon relative orbital
angular momentum.} the total angular
momentum of the final vector particle  $j_{V}$ and of the initial photon
$j_{\gamma}$ with respect to the nucleon; and the multipole type of the photon
$\phi_{\gamma}.$ Here,  $\phi_{\gamma}$ denotes either electric
$\phi_{\gamma}=E(j_{\gamma}=\ell_{\gamma}\pm 1)$ or magnetic
$\phi_{\gamma}=M(j_{\gamma}=\ell_{\gamma})$ multipole type.

Introducing complete sets of initial and final angular momentum states, the
$T$ matrix element can be brought into the following form:
\begin{eqnarray}
<\vec{q}\,\lambda_{V}\,m_2 \mid &T& \mid \vec{k}\,\lambda_{\gamma}\,m_1 >
\nonumber\\
&&= \sum_{\alpha} <m_2 \mid {\cal P}^{\lambda_{V}
\lambda_{\gamma}}_{\alpha} \mid m_1 >T_{\alpha}\ .\label{eq1}
\end{eqnarray} In other words:
\begin{equation}
{\cal J}_{\lambda_{V}\lambda_{\gamma}}=\sum_{\alpha}
{\cal P}^{\lambda_V\lambda_{\gamma}}_{\alpha}\ T_{\alpha}\ ,
\end{equation}
\noindent
where $\alpha\equiv(J,j_{V},\ell_{V},j_{\gamma},\phi_{\gamma}).$
The multipole amplitude $T_{\alpha}$ and the transformation operator
${\cal P}^{\lambda_V\lambda_{\gamma}}_{\alpha}$ are defined as

\begin{equation}
T_{\alpha}\equiv
<(\ell_{V}1)(j_{V}\frac{1}{2})\,JM \mid T \mid (j_{\gamma}\frac{1}{2})\,
\phi_{\gamma}\,JM>,
\end{equation} and
\begin{eqnarray}
&&<m_2 \mid {\cal P}^{\lambda_{V}\lambda_{\gamma}}_{\alpha} \mid m_1 >
\equiv \sum_M < \vec{q}\,\lambda_V\,m_2\mid q\,(\ell_V1)\,(j_V\frac{1}{2})\,JM>
<k\,\phi_{\gamma}\,(j_{\gamma}\frac{1}{2})\,JM \mid
\vec{k}\,\lambda_{\gamma}\,m_1>.
\end{eqnarray} Note that $T_{\alpha}$ is independent of $M.$
As seen above, the  quantum numbers describing the initial and final states
are quite similar, which is natural since both contain a vector
particle($\gamma$ or $V$) and a nucleon. The only asymmetry in the quantum
numbers is the use of $\ell_{V}$ in the final state instead of photon
multipole type $\phi_{\gamma}.$\footnote{An alternative definition for
multipole amplitudes is given in Ref\cite{alternative}, where not only
the initial photon but also the final vector meson is described by multipole
quantum numbers. In their approach, instead of using the final state orbital
angular momentum they refer to the multipolarity of the vector meson using a
second set of $(E,M,L)$ labels. For purposes of truncating near threshold and
identifying resonances, it is more convenient to keep the final state orbital
angular momentum as a good quantum number. In addition, our definition is a
natural generalization of the pseudoscalar meson case.}
Coupling of angular momenta, which is implied
by the use of parenthesis in the above expression, is pictured in
Figure~\ref{addit}:
\begin{figure}[thb]
\psfig{figure=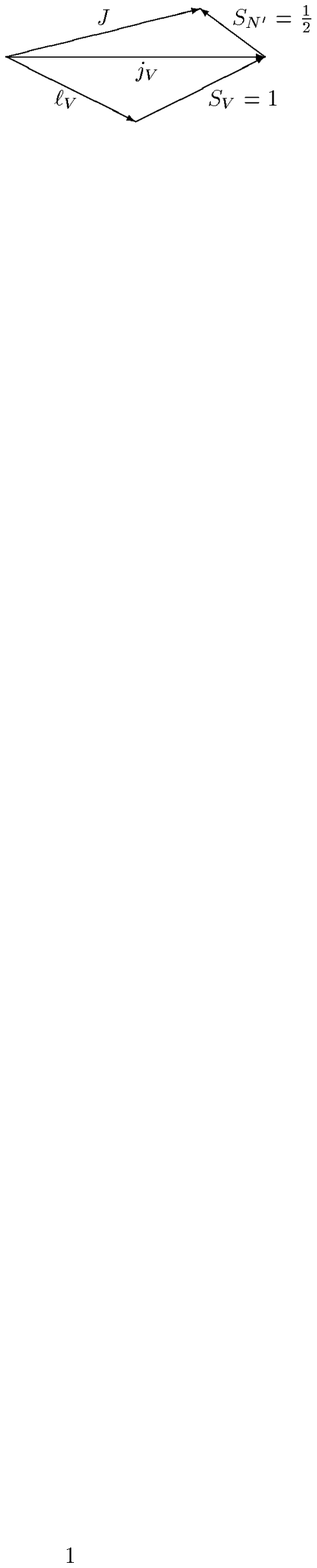}
\caption{Final State Angular Momenta Addition}
\label{addit}
\end{figure}

A plane wave photon state $|\vec{k}\, \lambda_{\gamma}>$ is characterized
by its momentum
$\vec{k}$ and helicity $\lambda_{\gamma}.$ A multipole photon state
$|j_{\gamma}\,
 m_{\gamma}, \phi_{\gamma} >$ is characterized by its total angular momentum
$j_{\gamma}$ and its projection onto a fixed $\hat{z}$ axis, $m_{\gamma},$ and
the multipole type $\phi_{\gamma}.$
The overlap of a photon plane wave state of definite helicity with a photon
multipole state is given by:

\begin{equation}
<j_{\gamma}\,m_{\gamma}\,\phi_{\gamma} \mid \vec{k}\,\lambda_{\gamma} >
 = -\xi_{\phi_{\gamma}}\sqrt{\frac{2j_{\gamma}+1}{8\pi}}\,
{\cal D}^{(j_{\gamma})}_{m_{\gamma}\lambda_{\gamma}}(\hat{k}),
\end{equation}

\noindent
where $\xi_{\phi_{\gamma}}$ is $1(\lambda_{\gamma})$ for Electric(Magnetic)
multipoles, respectively. (We follow the phase convention of Ref~\cite{GW} for
the photon multipole state.)
Using this expression, matrix elements of the
transformation operator ${\cal P}$ can be simplified as:

\begin{eqnarray}
&&<m_2 \mid {\cal P}^{\lambda_{V}\lambda_{\gamma}}_{\alpha} \mid m_1 > =
 C^{\lambda_{V}\lambda_{\gamma}}_{\alpha}(-1)^{\lambda_{V}}\sum_{m_{\gamma}
m_{V}M}{\cal D}^{(j_{V})*}_{m_{V}\lambda_{V}}(\hat{q})\nonumber\\
&&\times {\cal D}^{(j_{\gamma})}_{m_{\gamma}\lambda_{\gamma}}(\hat{k})
\left(\begin{array}{ccc}j_{V}&1/2&J\\ m_{V}&m_2&-M\end{array}\right)
\left(\begin{array}{ccc}j_{\gamma}&1/2&J\\m_{\gamma}&m_1&-M\end{array}\right).
\end{eqnarray}

\noindent
All angular momentum constraints are stored in the Wigner ${\cal D}$'s and in
the three-$j$ symbols. In the above expression, $C$ is given by:

\begin{eqnarray}
&&C^{\lambda_{V}\lambda_{\gamma}}_{\alpha} = \frac{(-1)^{j_{\gamma}-
\lambda_{\gamma}}}{2\sqrt{2}\pi}\xi_{\phi_{\gamma}}(J+\frac{1}{2})
 \nonumber \\
&&\times\sqrt{(2j_{V}+1)(2\ell_{V}+1)(2j_{\gamma}+1)}\left(
\begin{array}
{ccc}\ell_{V}&1&j_{V}\\ 0&-\lambda_{V}&\lambda_{V} \end{array}\right).
\end{eqnarray}

\subsection{Helicity and multipole amplitudes}
In order to express the multipole amplitudes in terms of the helicity
amplitudes, we need to invert Eq.~(\ref{eq1}). Therefore, an orthogonality
relationship for the transformation operator ${\cal P}$ is needed.
The operator ${\cal P}$ obeys the following orthogonality property:

\begin{eqnarray}
&&\int\int d\Omega_{\hat{q}}\,d\Omega_{\hat{k}}\sum_{m_1m_2
\lambda_{\gamma}\lambda_{V}} <m_2 \mid {\cal P}^{\dag \lambda_{V}
\lambda_{\gamma}}_{\alpha} \mid m_1 >\nonumber\\
&&\times < m_1 \mid {\cal P}^{\lambda_{V}\lambda_{\gamma}}_{\alpha'}
 \mid m_2 >=(2J+1)\delta_{\alpha\alpha'}.\label{ortho}
\end{eqnarray}
\noindent
Using this property, Eq.~(\ref{eq1}) can be inverted to yield the multipole
amplitudes:

\begin{eqnarray}
&&T_{\alpha}=\frac{1}{2J+1}\int\int d\Omega_{\hat{q}}\,d\Omega_{\hat{k}}
\sum_{m_1m_2\lambda_{\gamma}\lambda_{V}}\nonumber\\
&&\times <m_1 \mid {\cal P}^{\dag \lambda_{V}\lambda_{\gamma}}_{\alpha}
\mid m_2 ><\vec{q} \lambda_{V}m_2 \mid T \mid \vec{k}\lambda_{\gamma}m_1>.\
\label{invert}
\end{eqnarray}

\noindent
Equations~(\ref{ortho},\ref{invert}) remain valid when $m$'s(nucleon spin
projections along the $z$-axis) are replaced by the nucleon helicities
$\lambda$'s. Matrix elements of ${\cal P}$ in a nucleon helicity basis can  be
generated readily by Wigner rotations of the initial and final nucleon states.
Such nucleon helicity-based projection operators can be used to relate the
helicity and multipole amplitudes. The explicit expression is:

\begin{eqnarray}
&&<\lambda_{N'} \mid {\cal P}^{\lambda_{V}\lambda_{\gamma}}_{\alpha} \mid
\lambda_{N} > = C^{\lambda_{V}\lambda_{\gamma}}_{\alpha}\sum_{M's}
{\cal D}^{(J)*}_{M''M'}(\hat{q})\nonumber \\
&&\times {\cal D}^{(J)}_{M''M}(\hat{k})\left(\begin{array}{ccc}j_{V}&1/2&J\\
\lambda_{V}&-\lambda_{N'}&-M'\end{array}\right)\left(
\begin{array}{ccc}
j_{\gamma}&1/2&J\\ \lambda_{\gamma}&-\lambda_{N}&-M\end{array}\right).
\end{eqnarray}
\noindent
As a result, it is found that the multipole amplitudes
 are directly related to
the partial wave helicity amplitudes $<\lambda_{V}\lambda_{N'}
\mid T^J \mid \lambda_{\gamma}\lambda_N >$ by:
\begin{eqnarray}
&&T_{\alpha}=\frac{4\pi}{2J+1}\sum_{\lambda_{\gamma}\lambda_{V}
\lambda_N\lambda_{N'}}C^{\lambda_{V}\lambda_{\gamma}}_{\alpha}
\left(
\begin{array}{ccc}
j_{V}&1/2&J\\\lambda_{V}&-\lambda_{N'}&-\Lambda_{f}
\end{array}\right)\nonumber\\
&&\times \left(
\begin{array}
{ccc}j_{\gamma}&1/2&J\\\lambda_{\gamma}&-\lambda_N&-\Lambda_i\end{array}\right)
<\lambda_{V}\lambda_{N'} \mid T^J \mid \lambda_{\gamma}\lambda_N >.
\label{multipole}
\end{eqnarray}

Although $T$ is labeled by five quantum numbers, $j_{\gamma}$ is dictated by
parity, once the other quantum numbers are specified. This property follows
since the
initial and final state parities are respectively $-(-1)^{j_{\gamma}+\Phi}$
and $-(-1)^{\ell_{V}}$ where\footnote{Extra minus signs are due to the
intrinsic parity of vector particles.} $\Phi$ is 0(1) for Magnetic(Electric)
multipoles, or from Eq.~(\ref{multipole}) it is seen that

\begin{equation}
T_{\alpha}\sim 1+(-1)^{\ell_{V}+j_{\gamma}+\Phi}.\label{prop}
\end{equation}
\noindent
This result guarantees that there are at most 12 nonzero multipole amplitudes
for any $J.$ For a given $J,$ there may be at most two different
$j_{V}$'s $(J\pm\frac{1}{2}),$ three different $l_{V}$'s $(j_{V},j_{V}
\pm 1),$ two different $j_{\gamma}$'s $(J\pm\frac{1}{2})$ and two
multipole types $\Phi=0,1.$ Thus, the total number of possible multipole
amplitudes is $2\times 3\times 2\times 2=24.$ Looking at
the Eq.~(\ref{prop}), one sees that for each set of quantum numbers either
$\Phi=0$ or $\Phi=1$ gives a nonzero multipole amplitude. Therefore, one
deduces that there are 12 complex multipole amplitudes for vector meson
photoproduction. On the other hand, there are 24 complex helicity matrix
elements. Since they are interrelated by parity invariance, only 12 helicity
amplitudes are linearly independent.
\\
\subsection{Labeling of multipole amplitudes}

We now define multipole amplitudes for vector meson photoproduction by
generalizing
the notation used for pseudoscalar meson production.
 As in the pseudoscalar meson case, the labels $E,M$  are used to denote
electric and magnetic multipoles. But for vector meson photoproduction, an
additional $\pm,0$ designation is  used to indicate how the vector meson's
orbital angular momentum is added to its spin. The generalization of the
multipoles notation to the vector meson case is therefore:

\begin{eqnarray}
E^{2(J-\ell_{\pi})}_{\ell_{\pi}} &\longrightarrow&
 E^{2(J-j_{V})}_{\ell_{V},(j_{V}-\ell_{V})},\nonumber\\
M^{2(J-\ell_{\pi})}_{\ell_{\pi}} &
\longrightarrow& M^{2(J-j_{V})}_{\ell_{V},(j_{V}-\ell_{V})},\nonumber
\end{eqnarray}
\noindent
where the LHS applies to the pseudoscalar meson case and the RHS to the vector
meson production case. The superscript is stipulated by giving  the
sign of $2(J-\ell_{\pi})$(pseudoscalar meson) or   $2(J-j_{V})$(vector meson).
Thus if
$J=\ell_{\pi}\pm 1/2$ or $J=j_{V}\pm 1/2$, the superscript is $\pm.$
Also if $j_V=\ell_V\pm 1, j_V=\ell_V,$ the mutipole has a subscript label of
$\pm$ or $0,$ respectively.
 We define multipole
amplitudes for
vector meson photoproduction  in terms of the earlier $T-$matrix as:

\begin{eqnarray}
E^{2(J-j_{V})}_{\ell_{V},(j_{V}-\ell_{V})}&\equiv&
 \frac{T^{J}_{j_{V}\ell_{V}j_{\gamma}E}}{\sqrt{j_{\gamma}(j_{\gamma}+1)}},\\
M^{2(J-j_{V})}_{\ell_{V},(j_{V}-\ell_{V})}&
\equiv& \frac{T^{J}_{j_{V}\ell_{V}j_{\gamma}M}}
{\sqrt{j_{\gamma}(j_{\gamma}+1)}},
\end{eqnarray}
\noindent
with $\phi_{\gamma}=E$ and $\phi_{\gamma}=M.$
For example, a set of quantum numbers $J=3/2,$ $\ell_{V}=2,$
 $j_{V}=1$ and $\phi_{\gamma}=M$ corresponds to : $M^{+}_{2-},$ while
$j_{\gamma}$ is determined by parity invariance and the triangle inequality
$\Delta (J,j_{\gamma},\frac{1}{2}).$ Thus, for the example given above,
the initial and final state parities are respectively:
$-(-1)^{j_{\gamma}}$ and $-(-1)^{\ell_{V}}=-(-1)^2.$ Therefore, $j_{\gamma}$
has to be an even integer. At the same time, $j_{\gamma}$ has to satisfy the
triangle inequality $\Delta (\frac{3}{2},j_{\gamma},\frac{1}{2}),$ which
leaves only one option, $j_{\gamma}=2.$ All possible sets of quantum numbers
and the corresponding amplitudes are listed in Table~\ref{table1}. The above
example($M^{+}_{2-}$) appears in the 6th line of this table.
Some amplitudes are not physical;  namely, $E^{\pm}_{0-},$ $M^{\pm}_{0-},$
$E^-_{1-},$ $M^-_{1-},$  which violate the inequality $J\geq 1/2,$  and
$E^{\pm}_{00},$  $M^{\pm}_{00},$ which violate the triangle inequality
$\Delta (J_{V},\ell_{V},1),$  and $E^{-}_{10},$  $M^{-}_{0+},$ $E^{+}_{1-},$
$M^{-}_{2-},$ which violate the condition $j_{\gamma}\geq 1.$

Having related the helicity amplitudes to multipole
amplitudes, we analyze next the spin observables near
threshold, where the multipoles are most useful.

\section{THRESHOLD BEHAVIOR}
We base our analysis near threshold on the physical assumption that,
in the absence of special dynamics, multipole
amplitudes behave as $T\sim q^{\ell_{V}}$ for a low final state
momentum $\vec{q}.$  In addition to this assumption, the fact that quantum
numbers
have to satisfy certain triangle inequalities severely restricts the number of
multipole amplitudes. For example, for $J=\frac{1}{2}$
there are only 4 multipole amplitudes instead of 12; namely, $E^{-}_{0,+},$
$E^{-}_{2,-},$$M^{+}_{1,-}$ and $M^{-}_{1,0}.$
The maximum number of 12 amplitudes occurs
only  when $J\geq\frac{3}{2}.$
A list of these amplitudes, along with their relationship to partial wave
helicity amplitudes is presented in Appendix A.
In the following discussion,
spin observable ``profiles" are introduced:
\begin{equation}
C^{\gamma NN'V}_{l,k,k',ij} \equiv \frac{1}{4} {\rm Tr}[T(\sigma_{\gamma}^{l}
\sigma_{N}^{k})T^{\dag}(S^{ij}_V{\sigma}_{N'}^{k'})].
 \end{equation}  These ``profile" functions are determined by
bilinear products of amplitudes.  The spin observables, $\Omega,$ themselves
are defined by the ratio of the above profile functions
with the function
\begin{equation}
I(\theta) \equiv \frac{1}{4} {\rm Tr} [TT^{\dag}],
\end{equation} e.g.,
\begin{equation}
\Omega^{\gamma NN'V}_{l,k,k',ij} \equiv \frac{C^{\gamma NN'V}_{l,k,k',ij} }
 {I(\theta)  }.
\end{equation} In this paper,  we will only use the profile functions.

The above trace is over spin-space helicity quantum numbers $\lambda_{\gamma},
\lambda_N,\lambda_{V},\lambda_{N'}.$ Here $\sigma$'s are  $2\times 2$
Pauli spin matrices. The vector meson matrix $S^{ij}_{V}$ is a $3\times 3$
matrix where: $S^{00}_V\equiv I,$ $S^{0i}_V$ (i=1,2,3) are the usual spin-1
matrices $S^i,$\footnote{To permit up to 100\% polarization for a
vector meson in any direction, $S^i$ matrices need to be renormalized as
$\vec{{\cal S}}=\sqrt{\frac{3}{2}}\,\vec{S}.$ Therefore, $\vec{{\cal S}}
\cdot\vec{{\cal S}}=3,$ which guarantees that each component
is normalized to be
$\le 1.$} and five independent Cartesian tensor operators $S^{ij}_V$
$(i<j=1,2,3)$ are defined as:
\begin{eqnarray}
S^{ij}_V&=&\frac{1}{2}(S^iS^j+S^jS^i)-\frac{2}{3}\delta^{ij}.
\end{eqnarray}
For example, the triple spin observable which represents a linearly polarized
beam in the $y$ direction,\footnote{A discussion of photon polarization
and Stokes parameters is given in Ref.~\cite{TabakinFasanoSaghai}.}
a target polarized in the $y$ direction and a recoil vector meson polarized in
the $z'$ direction is described by $C^{\gamma NV}_{x,y,z'}.$ In this example,
the vector polarization of the vector meson is denoted by only $z'$ rather than
$0z'$. For simplicity, the vector polarization of the vector meson will be
labeled without the extra 0, while its tensor polarization is labeled
  by two indices.

\subsection{Truncation for $\ell_{V}=0$}
For $S-$waves $\ell_{V}=0,$ there are only three multipole amplitudes;
 $E^{-}_{0,+}, E^{+}_{0,+}, M^{+}_{0,+}.$ In contrast,
 for a pseudoscalar meson only
one multipole $E^{+}_{0}$ exists for $\ell_{\pi}=0.$ The differential cross
section is flat and given by:
\begin{equation}
\sigma(\theta)=\frac{q}{2k}\,I(\theta),
\label{cross1}
\end{equation}
where
\begin{equation}
I(\theta)=\vert E^{-}_{0,+}\vert^{2}+2\vert E^{+}_{0,+}
\vert^{2}+6\vert M^{+}_{0,+}\vert^{2}.
\label{cross2}
\end{equation}
At this level of truncation, all single spin observables vanish, except for
the tensor polarization of the vector meson. The angular dependencies of the
tensor polarization of the vector meson very near threshold are:
\begin{eqnarray*}
C^{V}_{x'x'}&=&(3x^2-2)\,A\\
C^{V}_{y'y'}&=&A\\
C^{V}_{z'z'}&=&(1-3x^2)\,A\\
C^{V}_{x'z'}&=&-3x\,\sin\theta\,A,
\end{eqnarray*}
\noindent
where $x=\cos\theta\equiv\hat{q}\cdot\hat{k},$ and $A,$  whose definition is
given in Appendix C, is a bilinear combination of multipole amplitudes.

Since all four vector meson tensor polarizations listed above involve the same
dynamical factor $A,$ they contain the same multipole information near
threshold.
Therefore, measurement of only one of 4 tensor polarizations is sufficient.
We list the multipole expansion of 4 tensor polarizations and 37 nonzero
double spin
observables using the $\ell_V=0$ truncation in Appendices B and C. Although
there are 42 nonzero observables(cross section + 4 single spin + 37 double
spin) near threshold, only 5 of them are necessary to determine three
multipole amplitudes with their relative phases\footnote{Since the overall
phase of amplitudes is arbitrary, there are only 5 numbers at each near
threshold energy: 3 magnitudes and 2 relative phases.}.
Therefore, three multipole amplitudes and two phases can be completely
determined by measuring  the cross section, a single spin observable
$C^{V}_{y'y'},$ and three double spin observables. {\it Therefore, even very
near threshold determination of the magnitudes of nonvanishing multipole
amplitudes $E^{-}_{0,+},E^{+}_{0,+},M^{+}_{0,+}$ require measurements of at
least one double spin observable.} In comparison, for the case of pseudoscalar
photoproduction one needs only the cross section to determine the magnitude of
$E_0^+$.  A full list of observables is provided in Appendices B \& C.
Examination of that list shows that many observables have common dynamical
factors, while their overall angular dependence differs. These experiments
can be thought of belonging to the same class. In choosing a set of
experiments, one needs only one experiment from a given class in order to
avoid redundant information.
In selecting experiments, one must of course take into account realistic
questions of feasibility and costs. A full analysis of which experiments are
needed to determine all 12 amplitudes is presented in
Ref.~\cite{PichowskySavkliTabakin},  where transversity amplitudes are shown
to be particularly advantageous.

\subsection{Nodal structure of helicity amplitudes
and observables near threshold}
The study of the nodal behavior of spin observables for vector mesons could be
a valuable tool in analyzing the underlying dynamics, as suggested in a
recent ${\rm K}^+$ photoproduction study(see Ref.~\cite{SaghaiTabakin}).
At this level of truncation($\ell_V=0$), helicity amplitudes
$H_{i,\lambda_{V}}(\theta)$\footnote{See Eq.~(\ref{labeling}) for the
definition of $H_{i,\lambda_{V}}(\theta).$}  are related to each other by the
following equations:
\begin{eqnarray*}
&&H_{1,-1}(\theta)=-H_{3,1}(\pi-\theta)\\
&&H_{1,0}(\theta)=-H_{3,0}(\pi-\theta)=\sqrt{2}H_{11}(\pi-\theta)\\
&&H_{1,0}(\theta)=-\sqrt{2}H_{3,-1}(\theta)\\
&&H_{2,0}(\theta)=H_{4,0}(\pi-\theta)\\
&&H_{2,-1}(\theta)=H_{4,1}(\pi-\theta)\\
&&H_{2,1}(\theta)=H_{4,-1}(\pi-\theta)
\end{eqnarray*}
Most of the 12 helicity amplitudes have only
endpoint($\theta=0^\circ,180^\circ$) nodes
near threshold. Exceptions to this are $H_{2,0},H_{2,1},H_{4,-1}$
and $H_{4,0},$ which may have 1 intermediate node depending on the relative
strength of the multipole amplitudes. As one goes to the $\ell_V=1$
truncation, the same four helicity amplitudes maintain their relatively rich
nodal structure, now with a possibility of having 2 intermediate nodes;
whereas, other helicity amplitudes may develop only 1 intermediate node. A
full list of these amplitudes, which can be used to construct spin observables
for $\ell_{V}\leq 1,$ is presented in Appendix A.
For $\ell_{V}=0,$ the nodal structure of single and double spin observables are
mostly due to the overall factors of sines and cosines. When one removes
these overall factors, all observables are flat, except the two
double spin observables; namely, $C^{N'V}_{x',x'}$ and $C^{N'V}_{z',z'}.$
These observables could have two intermediate($0<\theta<\pi$) nodes even
very near threshold, depending on the relative strength of the multipole
amplitudes. The angular behavior of observables at $\ell_{V}= 0$ resonances,
that is, when only one amplitude is nonzero,
is presented in Table~\ref{table2}.
 That table shows how such resonances might manifest themselves in the
angle dependence of spin observables in the absence of
 $\ell\geq 1$ multipoles.\\

Above the threshold region, multipoles with higher $\ell_V$ begin to
contribute. Although we have derived expressions for double and triple spin
observables, only the results of single spin observables are presented
here. One can easily reproduce any observable using the helicity
amplitudes given in Appendix A.4 for $\ell_{V}\leq 1.$ With the inclusion of
$\ell_{V}=1$ multipole amplitudes, single spin observables have the following
general structure at threshold:
\begin{eqnarray*}
C^{N}_{y}&=&[a_{N}+b_{N}\cos\theta]\sin\theta\\
C^{\gamma}_{x}&=&b_{\gamma}\sin^{2}\theta\\
C^{N'}_{y'}&=&[a_{N'}+b_{N'}\cos\theta]\sin\theta\\
C^{V}_{y'}&=&[a_{V}+b_{V}\cos\theta]\sin\theta
\end{eqnarray*}
where $a$ and $b$'s are real. Assuming a threshold $q^{\ell_V}$ dependence for
the onset of multipoles,
 we found that the $a$'s are of order $q$; whereas, $b$'s are of
order $q^2.$ Therefore, very near threshold the $a$'s will be dominant in
determining the angular dependence of the spin observable profiles:
\begin{eqnarray*}
C^{N}_{y}&=&a_{N}\sin\theta\\
C^{\gamma}_{x}&=&0\\
C^{N'}_{y'}&=&a_{N'}\sin\theta\\
C^{V}_{y'}&=&a_{V}\sin\theta.
\end{eqnarray*}
{}From the above angular dependence,  one sees that  near threshold these
profiles will have only endpoint nodes with
intervening nodes developing as the incident energy increases. Of course,
for nodes to develop it is required that $b\geq a.$
 The development of these nodes
 can serve to reveal important underlying dynamics
as has been illustrated recently
 in Ref.~\cite{SaghaiTabakin}.  Note that the above profile functions
correspond to the single spin observables
$T$(target polarization), $\Sigma$(photon asymmetry),
$P$(recoil baryon polarization) and $P_V$(vector meson polarization),
respectively~\cite{SaghaiTabakin}.
\section{CONCLUSION}
Spin observables are very sensitive probes of hadron structure. With
the construction of CEBAF, polarized beams and targets will be available for
high precision spin observable measurements. In this context, there is
renewed interest in the photoproduction of vector mesons. In this paper,
we derived exact expressions for the multipole amplitudes in terms of the
helicity amplitudes for all $J$ values. We found that there are only 4
multipole amplitudes for the $J=1/2$ case;
whereas, there are 12 multipole amplitudes for all other $J$'s. Then we
analyzed the observables for $\ell_V=0$ and $\ell_V\leq 1$ truncations. We
found that near threshold($\ell_V=0$), there are only 4 single spin
observables which are the tensor polarizations of vector meson. (Note that
there are no nonzero spin observables for photoproduction of pseudoscalar
mesons at that level of truncation.) For $\ell_V=0,$ there are 3 multipole
amplitudes, as opposed to 1 for the photoproduction of pseudoscalar mesons.
According to our results, at threshold only two double spin observables are
able to have intermediate nodes when one neglects overall angular factors.
These two observables are $C^{N'V}_{x',x'}(\theta)$ and
$C^{N'V}_{z',z'}(\theta).$ Similarly, only four Helicity
amplitudes($H_{2,0},H_{2,1},H_{4,-1},H_{4,0}$) are able to
have intermediate nodes when overall angular factors are not considered.

For a full determination of the 3 $\ell_V=0$
 multipole amplitudes with their relative phases,
a set of 5 observables, the cross section + single spin(vector meson tensor
polarization) + 3 double spin observables, are needed.
 Measurement of a double spin
observable(besides the cross section and vector meson tensor polarization) is
unavoidable even for a determination of just the magnitudes of
the three nonvanishing amplitudes $E^{-}_{0,+},E^{+}_{0,+},M^{+}_{0,+}$ near
threshold. This is to be compared to the case of photoproduction of
pseudoscalar mesons, where measurement of the cross section is sufficient
to determine the magnitude of $E^{+}_{0}$ near threshold.
 Although the number of nonvanishing amplitudes will increase with increasing
energy, threshold analysis provides a first approximation to the expected form
of observables near threshold. Since multipole amplitudes have definite
$\ell$-values, the presence of a resonance will be signaled by the large
contribution of a particular multipole amplitude. Although most of the
discussion has been based on an $\ell_{V}=0$ truncation, the helicity
amplitudes
for $\ell_{V}\leq 1,$ from which all spin observables can be generated, are
listed in Appendix A.4.

\acknowledgments
\c{C}. \c{S}avkl\i \
 and F.Tabakin are thankful to the Department of Physics of National Taiwan
University for their kind hospitality during visits there.
In addition, S. N. Yang thanks the Nuclear Theory Group at LBNL for their
 warm hospitality.
 This research was
supported, in part, by the National Science Council of ROC under grant
NSC82-0212-M002-170-Y and by the U.S. National Science Foundation INT-9021617.
One of the authors(\c{C}.~\c{S}.) has been supported by an Andrew Mellon
Predoctoral Fellowship.
\appendix

\section{Helicity Amplitudes and Multipoles}

In this Appendix, $J=\frac{1}{2}$ and $J=\frac{3}{2}$ multipole amplitudes are
presented, and $J=\frac{1}{2}$ partial
wave helicity amplitudes are expressed in terms of multipoles. Then the
helicity amplitudes in terms of multipoles for $\ell_{V}=0$ and
$\ell_{V}\leq 1$ are given.

Helicity amplitudes $H_{i\lambda_{V}}$ are defined  by:
\begin{eqnarray}
H_{1,\lambda_{V}}&\equiv& <\lambda_{V},\lambda_{N'}=+1/2 \mid
T  \mid \lambda_{\gamma}=1 ,\lambda_N=-1/2 >\\
H_{2,\lambda_{V}}&\equiv& <\lambda_{V},\lambda_{N'}=+1/2 \mid
T  \mid \lambda_{\gamma}=1 ,\lambda_N=+1/2 >\\
H_{3,\lambda_{V}}&\equiv& <\lambda_{V},\lambda_{N'}=-1/2 \mid
T  \mid \lambda_{\gamma}=1 ,\lambda_N=-1/2 >\\
H_{4,\lambda_{V}}&\equiv& <\lambda_{V},\lambda_{N'}=-1/2 \mid
T  \mid \lambda_{\gamma}=1 ,\lambda_N=+1/2 >.\label{labeling}
\end{eqnarray}
The partial wave helicity amplitudes $h^J_{i\lambda_{V}}$ are defined
similarly by
\begin{equation}
h^J_{i\lambda_{V}}\equiv <\lambda_{V},\lambda_{N'}\mid T^J \mid
\lambda_{\gamma},\lambda_N>,
\end{equation}
where the set of helicity quantum numbers for each $i$ label is the same as
in Eq.~(\ref{labeling}). Therefore, $h^J_{i\lambda_{V}}$ is related to
the helicity amplitude $H_{i\lambda_{V}}$ by:
\begin{equation}
H_{i,\lambda_{V}}=\sum_{JM}\frac{2J+1}{4\pi}
{\cal D}^{(J)*}_{M\Lambda_f}(\hat{q})\
{\cal D}^{(J)}_{M\Lambda_i}(\hat{k})\ h^{J}_{i\lambda_{V}},
\end{equation}
where $\Lambda_f=\lambda_{V}-\lambda_{N'}$ and
$\Lambda_i=\lambda_{\gamma}-\lambda_N.$ The index $i$ varies from 1 to 4.
The general expression for the partial wave amplitude is:
\begin{eqnarray}
h^J_{i,\lambda_V}&=&\frac{4\pi}
{2J+1}\sum_{j_V\ell_Vj_{\gamma}\phi_{\gamma}}
C^{J\lambda_V\lambda_{\gamma}}_{j_V\ell_Vj_{\gamma}
\phi_{\gamma}}T^J_{j_V\ell_Vj_{\gamma}\phi_{\gamma}}\nonumber\\
&&\times\left(\begin{array}{ccc}j_V&1/2&J\\\lambda_{V}&-\lambda_{N'}
&-\Lambda_f\end{array}\right)\left(\begin{array}{ccc}j_{\gamma}&1/2&J\\
\lambda_{\gamma}&-\lambda_N&-\Lambda_i\end{array}\right),\label{partialwave}
\end{eqnarray}
where $C^{J\lambda_V\lambda_{\gamma}}_{j_V\ell_Vj_{\gamma}\phi_{\gamma}}$
 and $T^J_{j_V\ell_Vj_{\gamma}\phi_{\gamma}}$ are given in the text.
\subsection{Partial wave helicity amplitudes versus multipoles for
 $J=\frac{1}{2}$}
For $J=1/2,$ there are four multipole amplitudes, which we collect
to form the following matrix
\begin{equation}
{\rm M}^{1/2}=[E^{-}_{0,+},E^{-}_{2,-},M^{+}_{1,-},M^{-}_{1,0}].
\end{equation}
The corresponding four helicity amplitudes also form a matrix defined by
\begin{equation}
{\rm H}^{1/2}=[h^{1/2}_{4,-1},h^{1/2}_{2,0},h^{1/2}_{4,0},h^{1/2}_{2,1}].
\end{equation}
Using Eq.~(\ref{partialwave}), we find the following
 matrix relationship between
 the partial wave helicity $H^{1/2}$ and multipole $M^{1/2}$ amplitudes
\begin{eqnarray}
\left[\begin{array}{c}h^{1/2}_{4,-1}\\\noalign{\medskip}{h^{1/2}_{2,0}}
\\\noalign{\medskip}{h^{1/2}_{4,0}}\\\noalign{\medskip}{h^{1/2}_{2,1}}\\
\end{array}
\right]&=&\frac{1}{\sqrt{6}}\left[
\begin{array}{cccc}\sqrt{2}&1&0&\sqrt{3}\\\noalign{\medskip}{1}
&-{\sqrt{2}}&-{\sqrt{3}}&0\\\noalign{\medskip}{1}&-{\sqrt{2}}&{\sqrt{3}}
&0\\\noalign{\medskip}{\sqrt{2}}&{1}&0&-{\sqrt{3}}\\\end{array}\right]
\left[\begin{array}{c}E^{-}_{0,+}\\\noalign{\medskip}{E^{-}_{2,-}}
\\\noalign{\medskip}{M^{+}_{1,-}}\\\noalign{\medskip}{M^{-}_{1,0}}
\end{array}\right]\nonumber
\end{eqnarray}

\widetext
\subsection{Partial wave helicity amplitudes versus multipoles for
 $J=\frac{3}{2}$}

Similarly, for $J=3/2,$ we have row matrices
\begin{eqnarray}
&&{\rm M}^{3/2}=[E_{0,+}^{+},E_{1,0}^{+},E_{2,-}^{+},E_{1,+}^{-}
,E_{2,0}^{-},E_{3,-}^{-},M_{0,+}^{+},M_{1,0}^{+},M_{2,-}^{+},
M_{1,+}^{-},M_{2,0}^{-},M_{3,-}^{-}],\\
&&{\rm H}^{3/2}=[h^{3/2}_{{1,-1}},h^{3/2}_{{2,-1}},h^{3/2}_{{3,-1}},
h^{3/2}_{{4,-1}},h^{3/2}_{{1,0}},h^{3/2}_{{2,0}},h^{3/2}_{{3,0}},
h^{3/2}_{{4,0}},h^{3/2}_{{1,1}},h^{3/2}_{{2,1}},
h^{3/2}_{{3,1}},h^{3/2}_{{4,1}}],
\end{eqnarray} which consist of 12 multipole and 12 helicity amplitudes. Using
 Eq.~(\ref{partialwave}), one can easily produce the linear relationship
 between these two sets of amplitudes, which involves a
 cumbersome $12\times 12$ matrix.

\mediumtext
\subsection{Helicity Amplitudes expanded in multipole basis for
 $\ell_{V}= 0$ truncation}
\mediumtext
Instead of partial waves, we now present the full
 amplitudes(see Eq.~(\ref{labeling})).
 The helicity amplitudes $H_{i\lambda_{V}}$ in terms of
multipoles for $\ell_{V}=0$ are given by:
\begin{eqnarray}
H^{\ell_{V}= 0}_{1,-1}&=&\frac{\sqrt{3}}{2}(-E^{+}_{0,+}
+M^{+}_{0,+})\sin(\frac{\theta}{2})(1-x),\\
H^{\ell_{V}= 0}_{1,0}&=&\frac{\sqrt{3}}{\sqrt{2}}
(-E^{+}_{0,+}+M^{+}_{0,+})
\cos(\frac {\theta}{2})(1-x) ,\\
H^{\ell_{V}= 0}_{1,1}&=&\frac{\sqrt{3}}{2}(-E^{+}_{0,+}
+M^{+}_{0,+})\sin(\frac
{\theta}{2})(1+x),\\
H^{\ell_{V}= 0}_{2,-1}&=&\frac{\sqrt{3}}{2}(E^{+}_{0,+}
+3\,M^{+}_{0,+})\cos(\frac
{\theta}{2})(1-x),\\
H^{\ell_{V}= 0}_{2,0}&=&\frac{1}{\sqrt{6}}\big{\{}2\,
E^{-}_{0,+}+E^{+}_{0,+}+3\,
M^{+}_{0,+}+3x\,(E^{+}_{0,+}+3\,M^{+}_{0,+})\big{\}}
\sin(\frac{\theta}{2}),\\
H^{\ell_{V}= 0}_{2,1}&=&\ \ \frac{\sqrt {3}}{6}\big{\{}4\
,E^{-}_{0,+}-E^{+}_{0,+}-3
\,M^{+}_{0,+}+3x\,(E^{+}_{0,+}+3\,M^{+}_{0,+})\big{\}}
\cos(\frac{\theta}{2}),\\
H^{\ell_{V}= 0}_{3,-1}&=&-\frac{\sqrt{3}}{2}(-E^{+}_{0,+}+M^{+}_{0,+})
\cos(\frac
 {\theta}{2})(1-x),\\
H^{\ell_{V}= 0}_{3,0}&=&-\frac{\sqrt{3}}{\sqrt {2}}(-E^{+}_{0,+}+M^{+}_{0,+})
\sin(\frac {\theta}{2})(1+x),\\
H^{\ell_{V}= 0}_{3,1}&=&-\frac{\sqrt{3}}{2}(-E^{+}_{0,+}+M^{+}_{0,+})
\cos(\frac
{\theta}{2})(1+x),\\
H^{\ell_{V}= 0}_{4,-1}&=&-\frac{\sqrt{3}}{6}\big{\{}-4\
,E^{-}_{0,+}+E^{+}_{0,+}
+3\,M^{+}_{0,+}+3x\,(E^{+}_{0,+}+3\,M^{+}_{0,+})\big{\}}
\sin(\frac{\theta}{2}),\\
H^{\ell_{V}= 0}_{4,0}&=&-\frac{1}{\sqrt{6}}\big{\{}-2\
,E^{-}_{0,+}-E^{+}_{0,+}
-3\,M^{+}_{0,+}+3x\,(E^{+}_{0,+}+3\,M^{+}_{0,+})\big{\}}
\cos(\frac{\theta}{2}),\\
H^{\ell_{V}= 0}_{4,1}&=&\ \ \frac{\sqrt{3}}{2}(E^{+}_{0,+}
+3\,M^{+}_{0,+})
\sin(\frac {\theta}{2})(1+x).
\end{eqnarray}
These are full $H(\theta)$ amplitudes; the $\ell_{V}= 0$
superscript just indicates
an $S-$wave truncation.
\widetext
\subsection{Helicity Amplitudes for $\ell_{V}\leq 1.$}
The results for $\ell_{V}\leq 1$ are obtained by adding the
following $\ell_{V}= 1$ helicity amplitudes to $\ell_{V}= 0$ terms
listed above
\begin{eqnarray}
H^{\ell_{V}= 1}_{1,-1}&=&\sin(\frac{\theta}{2})(1-x)\Big{\{}
E^{-}_{1,+}-M^{-}_{1,+}+\sqrt{5}(E^{+}_{1,0}-M^{+}_{1,0})+6(M^{+}_{1,+}
-E^{+}_{1,+})+10x(M^{+}_{1,+}-E^{+}_{1,+})\Big{\}}\frac{3}{\sqrt{40}},\\
H^{\ell_{V}= 1}_{1,0}&=&\cos(\frac{\theta}{2})(1-x)\Big{\{}E^{-}_{1,+}-M^{-}
_{1,+}+M^{+}_{1,+}-E^{+}_{1,+}+5x(M^{+}_{1,+}-E^{+}_{1,+})
\Big{\}}\frac{3}{\sqrt{5}},\\
H^{\ell_{V}= 1}_{1, 1}&=& \sin(\frac{\theta}{2})(1+x)\Big{\{}
E^{+}_{1,+}-M^{+}
_{1,+}+\frac{\sqrt{5}}{2}(M^{+}_{1,0}-E^{+}_{1,0})+\frac{3}{2}
(E^{-}_{1,+}-M^{-}
_{1,+})+10x(M^{+}_{1,+}-E^{+}_{1,+})\Big{\}}\frac{3}{\sqrt{40}},\\
H^{\ell_{V}= 1}_{2, -1}&=&\cos(\frac{\theta}{2})(1-x)\Big{\{}
\sqrt{5}(3E^{+}
_{1,0}+M^{+}_{1,0})+3E^{-}_{1,+}+M^{-}_{1,+}+2(2M^{+}_{1,+}
+E^{+}_{1,+})+10x(2M^{+}
_{1,+}+E^{+}_{1,+})\Big{\}}\frac{3}{\sqrt{40}},\\
H^{\ell_{V}= 1}_{2,0}&=&\sin(\frac{\theta}{2})\Big{\{}
-10\sqrt{2}M^{+}_{1,-}+3\sqrt{5}(2E^{-}
_{1,+}+M^{-}_{1,+})-6\sqrt{5}(E^{+}_{1,+}+2M^{+}_{1,+}),\nonumber\\
&&+x\big{[}6\sqrt{5}(M^{-}_{1,+}+3E^{-}_{1,+}+4M^{+}_{1,+}
+2E^{+}_{1,+})\big{]}+x^{2}
\big{[}30\sqrt{5}(E^{+}_{1,+}+2M^{+}_{1,+})\big{]}\Big{\}}
\frac{1}{10},\\
H^{\ell_{V}= 1}_{2,1}&=&\cos(\frac{\theta}{2})\Big{\{}
-\frac{4\sqrt{5}}{3}M^{-}_{1,0}
-(M^{-}_{1,+}+3E^{-}_{1,+})-2(E^{+}_{1,+}+2M^{+}_{1,+})+\frac{\sqrt{5}}{3}
(M^{+}_{1,0}+3E^{+}_{1,0}),\nonumber\\
&&+x\big{[}3(M^{-}_{1,+}+3E^{-}_{1,+})-\sqrt{5}(M^{+}_{1,0}+3E^{+}_{1,0})
-4(E^{+}_{1,+}+2M^{+}_{1,+})\big{]}+x^{2}10(E^{+}_{1,+}+2M^{+}_{1,+})
\Big{\}}\frac{3}{\sqrt{40}},\\
H^{\ell_{V}= 1}_{3,-1}&=&\cos(\frac{\theta}{2})(-1+x)\Big{\{}\sqrt{5}
(E^{+}_{1,0}-M^{+}_{1,0})+3(M^{-}_{1,+}-E^{-}_{1,+})+2(M^{+}_{1,+}
-E^{+}_{1,+})+10x(M^{+}_{1,+}-E^{+}_{1,+})\Big{\}}\frac{3}{\sqrt{40}},\\
H^{\ell_{V}= 1}_{3,0}&=&-\sin(\frac{\theta}{2})(1+x)\Big{\{}E^{+}_{1,+}
-M^{+}_{1,+}-E^{-}_{1,+}+M^{-}_{1,+}+5x(M^{+}_{1,+}
-E^{+}_{1,+})\Big{\}}6\sqrt{5},\\
H^{\ell_{V}= 1}_{3,1}&=&-\cos(\frac{\theta}{2})(1+x)\Big{\{}6(E^{+}_{1,+}
-M^{+}_{1,+})+M^{-}_{1,+}-E^{-}_{1,+}+\sqrt{5}(M^{+}_{1,0}-E^{+}_{1,0})
+10x(M^{+}_{1,+}-E^{+}_{1,+})\Big{\}}\frac{3}{\sqrt{40}},\\
H^{\ell_{V}= 1}_{4,-1}&=&-\sin(\frac{\theta}{2})\Big{\{}-6(E^{+}_{1,+}
+2M^{+}_{1,+})+\sqrt{5}(M^{+}_{1,0}+3E^{+}_{1,0})-3(M^{-}_{1,+}
+3E^{-}_{1,+})
-4\sqrt{5}(M^{-}_{1,0}+\frac{\sqrt{6}}{3}E^{-}_{0,1}),\nonumber\\
&&+3x\big{[}\sqrt{5}(M^{+}_{1,0}+3E^{+}_{1,0})-3(E^{-}_{1,+}+M^{-}_{1,+})
+4(E^{+}_{1,+}+2M^{+}_{1,+})\big{]}+30x^{2}(E^{+}_{1,+}+2M^{+}_{1,+})
\Big{\}}\frac{1}{\sqrt{40}},\\
H^{\ell_{V}= 1}_{4,0}&=&-\cos(\frac{\theta}{2})\Big{\{}M^{-}_{1,+}
+3E^{-}_{1,+}-3(E^{+}_{1,+}+2M^{+}_{1,+})-\sqrt{10}M^{+}_{1,-},\nonumber\\
&&+x\big{[}-3(M^{-}_{1,+}+3E^{-}_{1,+})-6(E^{+}_{1,+}+2M^{+}_{1,+})\big{]}
+15x^{2}(E^{+}_{1,+}+2M^{+}_{1,+})\Big{\}}\frac{1}{\sqrt{5}},\\
H^{\ell_{V}= 1}_{4,1}&=&\sin(\frac{\theta}{2})(1+x)\Big{\{}-M^{-}_{1,+}
-3E^{-}_{1,+}-2(E^{+}_{1,+}+2M^{+}_{1,+})-\sqrt{5}(M^{+}_{1,0}+3E^{+}_{1,0})
+10x(E^{+}_{1,+}+2M^{+}_{1,+})\Big{\}}\frac{3}{\sqrt{40}}.
\end{eqnarray}
Hence, adding $H^{\ell_{V}= 0}_{i,\lambda_V}$ to
$H^{\ell_{V}= 1}_{i,\lambda_V}$ yields all 12 helicity
amplitudes in terms of the $S$ and $P$ wave multipoles. These expressions are
useful for determining the energy evolution and nodal structure of observables
near threshold.
\widetext
\section{Single spin observables for $\ell_{V}= 0$}
In the following list of spin observables, various bilinear combinations of
multipole amplitudes are denoted by $A,B,\cdots$ etc. Their definitions
appear after the list of observables(e.g., profile functions).
Sets of observables which contain the bilinear combinations
  provide equivalent information. Only those observables which do not
vanish for $\ell_{V}= 0$ are presented below.

{\bf 1. Cross section}

 The cross section,  which we count as a single spin observable
was already presented in the text, see Eqs.~\ref{cross1},~\ref{cross2}.
For the pseudoscalar case the $S-$wave cross section
is simply
$
\sigma(\theta) = \frac{q}{2k}\,\vert E^{+}_{0}\vert^{2}.
$

{\bf 2. Tensor polarizations of vector meson}

Only four single spin observables are possibly nonzero for pure S-wave
multipoles, namely,
\begin{eqnarray*}
& C^{V}_{x'x'}=(3\cos^{2}\theta-2)\,A,\ \ \
& C^{V}_{y'y'}=A,\\
& C^{V}_{z'z'}=(1-3\cos^{2}\theta )\,A,\ \ \
& C^{V}_{x'z'}=-3 \cos\theta \sin\theta\,A.
\end{eqnarray*}
Since only one dynamical factor A appears, only one single spin
observable needs to be measured near threshold.
All other single spin observables vanish for $\ell_{V}= 0$ truncation.

\section{Double spin observables for $\ell_{V}=0$}

Double spin observables fall into six categories. Here the nonzero double spin
observables for each category are presented.

\noindent{\bf Beam-Target}

\begin{eqnarray*}
&&C^{\gamma N}_{z,z}=K.
\end{eqnarray*}

\noindent{\bf Target-Recoil}

The target-recoil observables depend only on the
two dynamical factors $B$ and $\frac{1}{3}\,I-4\,A,$ thus only two of the
following are independent.
\begin{eqnarray*}
C^{NN'}_{x,x'}=-\cos\theta\,B\ \ \  & &
 C^{NN'}_{x,z'}=\sin\theta\,B \\
&C^{NN'}_{y,y'}=-B &\\
C^{NN'}_{z,x'}=\sin\theta\,(\frac{1}{3}\,I-4\,A)\ \ \  &&
C^{NN'}_{z,z'}=\cos\theta\,(\frac{1}{3}\,I-4\,A)
\end{eqnarray*}

\noindent{\bf Beam-Recoil}

The beam-recoil observables depend only on one
 dynamical factor  $\frac{2}{3}\,I+4\,A+K$
\begin{eqnarray*}
& C^{\gamma N'}_{z,x'}=\sin\theta\,(\frac{2}{3}\,I+4\,A+K)\ \ \ \ \ \
&C^{\gamma N'}_{z,z'}=\cos\theta\,(\frac{2}{3}\,I+4\,A+K)
\end{eqnarray*}

\noindent{\bf Beam-Vector Meson}

The beam-vector meson observables,  which involve the vector meson
polarization depend only on one
 dynamical factor  $\frac{2}{3}\,I-2\,A$
\begin{eqnarray*}
&C^{\gamma V}_{z,x'}=\sin\theta\,(\frac{2}{3}\,I-2\,A)\ \ \ \ \ \
&C^{\gamma V}_{z,z'}=\cos\theta\,(\frac{2}{3}\,I-2\,A),
\end{eqnarray*}
\widetext
\noindent while those that involve the tensor polarization
depend only on $B,$ which already appeared in the target-recoil observables:
\begin{eqnarray*}
C^{\gamma V}_{x,x'z'}=\cos\theta \sin\theta\,
\displaystyle{\frac{B}{2}} \ \ \  &
C^{\gamma V}_{x,y'y'}= \,\displaystyle{\frac{B}{2}} &
C^{\gamma V}_{x,z'z'}=-\sin^{2}\theta\,\displaystyle{\frac{B}{2}}  \\
C^{\gamma V}_{y,x'y'}=-\cos\theta\, \displaystyle{\frac{B}{2}}\ \ \ \ &
C^{\gamma V}_{y,y'z'}= \sin\theta\, \displaystyle{\frac{B}{2}}.
\end{eqnarray*}
Thus,  near threshold it is not necessary to observe these tensor polarization
observables since the same factor $B$ appears in  the
target-recoil observables, which are perhaps easier to measure.

\noindent{\bf Target-Vector Meson}

For the following target-vector meson observables one needs to
 have a polarized target and measure the vector meson's polarization for two
different cases,  one depending on $F,$ the other on $I+6\,A+3\,K.$
\begin{eqnarray*}
C^{NV}_{x,x'}=\cos\theta\,F\ \  & &
C^{NV}_{x,z'}=\ \ -\sin\theta\,F\\
& C^{NV}_{y,y'}=F  &\\
C^{NV}_{z,x'}=\frac{1}{3}\,(I+6\,A+3\,K)\,\sin\theta\ \  & &
C^{NV}_{z,z'}=\frac{1}{3}\,(I+6\,A+3\,K)\,\cos\theta,
\end{eqnarray*}
For the tensor polarization observable near threshold,
all of the following depend on one dynamical factor $H.$
\begin{eqnarray*}
C^{NV}_{x,x'y'}=\sin\theta\,H\ \ \ \  &
C^{NV}_{x,y'z'}=\cos\theta \,H  & \\
C^{NV}_{y,x'x'}=-\cos\theta \sin\theta\,H\ \ \ \  &
C^{NV}_{y,x'z'}=(2\cos^{2}\theta-1)\,H \ \ \ \  &
C^{NV}_{y,z'z'}=\cos\theta \sin\theta\,H.
\end{eqnarray*}

\noindent{\bf Recoil-Vector Meson}

The dynamical factors $V,O$ and $S$ appear for
the polarization of the vector meson cases:
\begin{eqnarray*}
C^{N'V}_{x',x'}=\frac{1}{3}\,(I+6\,A+3\,K)-\cos^{2}\theta\,O\ \  & &
C^{N'V}_{x',z'}=\cos\theta\sin\theta\,O\\
&C^{N'V}_{y',y'}=\frac{1}{3}\,(I+6\,A+3\,K)-O& \\
C^{N'V}_{z',x'}=\cos\theta\sin\theta\,O\ \  & &
C^{N'V}_{z',z'}=S+\cos^{2}\theta\,O.
\end{eqnarray*}
Finally a single dynamic factor $R$ appears in
all recoil-baryon and vector meson tensor polarization cases:
\begin{eqnarray*}
C^{N'V}_{x',y'z'}= -\cos^{2}\theta\,R\ \ \ &
C^{N'V}_{x',x'y'}=-\cos\theta\sin\theta\,R & \\
C^{N'V}_{y',x'x'}= \cos\theta\sin\theta\,R \ \ \ &
C^{N'V}_{y',x'z'}=(2\cos^{2}\theta-1)\,R\ \ \ &
C^{N'V}_{y',z'z'}= -\cos\theta \sin\theta\,R  \\
C^{N'V}_{z',x'y'}= \sin^{2}\theta\,R\ \ &
C^{N'V}_{z',y'z'}= \cos\theta \sin\theta\,R.&
\end{eqnarray*}

In the above expressions the following dynamic combinations of the multipoles
appear:
\begin{eqnarray}
I&=&\alpha^2+2\,\beta^2+6\,\gamma^2\\
A&=&-\frac{1}{6}\,\beta^2+\frac{1}{2}\,\gamma^2-\frac{1}{3}\,\alpha\beta\,
\cos\phi_E-\alpha\gamma\,\cos\phi_M+\beta\gamma\,\cos(\phi_M-\phi_E)\\
B&=&-\beta^2+3\,\gamma^2-2\,\alpha\beta\,\cos\phi_E+2\,\alpha\gamma\,
\cos\phi_M-2\,\beta\gamma\,\cos(\phi_M-\phi_E)\\
F&=&\beta^2-3\,\gamma^2-\alpha\beta\,\cos\phi_E+\alpha\gamma\,\cos\phi_M+2\,
\beta\gamma\,\cos(\phi_M-\phi_E)\\
H&=&\frac{1}{2}\,\alpha\beta\,\sin\phi_E-\frac{1}{2}\,\alpha\gamma\,
\sin\phi_M+2\,\beta\gamma\,\sin(\phi_M-\phi_E)\\
K&=&-\alpha^2+\beta^2-3\,\gamma^2-6\,\beta\gamma\,\cos(\phi_M-\phi_E)\\
O&=&\beta^2-3\,\gamma^2-\alpha\beta\,\cos\phi_E-3\,\alpha\gamma\,
\cos\phi_M-6\,\beta\gamma\,\cos(\phi_M-\phi_E)\\
R&=&-\alpha\beta\,\sin\phi_E-3\,\alpha\gamma\,\sin\phi_M,
\end{eqnarray} where the magnitudes
 $\alpha,\beta,\gamma$ of the multipole amplitudes, and
two relative phases $\phi_E$ and $\phi_M$ are defined by
\begin{eqnarray}
E^{-}_{0,+}=\alpha,\ \ E^{+}_{0,+}=\beta\,e^{i\phi_E},\ \
M^{+}_{0,+}=\gamma\,e^{i\phi_M}.
\end{eqnarray} Since there are 3 S-wave multipoles,
 there are only 3 magnitudes and 2
independent phases to be determined near threshold. That implies doing 5
experiments. One can use the above expressions to select experiments that give
nonredundant multipole amplitude information. It should be clear that one
should choose only one experiment from a class of experiments with the same
dynamical coefficient in order to avoid redundancy. Since there are only 5
unknown functions, the above 8 relations are not independent.

\mediumtext
\begin{table}
\caption{ Quantum numbers and states for vector meson photoproduction}~
\tablenotetext[1]{Unphysical amplitudes (violation) :
 $E_{0-}^{\pm},M_{0-}^{\pm},E_{1-}^{-},M_{1-}^{-}$ ($J\ge 1/2$),\ \
$E_{00}^{\pm},M_{00}^{\pm}$ ($\Delta(j_{V},L,1)$),\
\ $E_{10}^{-},M_{0+}^{-},E_{1-}^{+},M_{2-}^{-}$ ($J_{\gamma}\ge 1$).
 Here $L$ denotes the vector meson-nucleon relative orbital angular momentum.}
\begin{tabular}{cccccc}
\multicolumn{2}{c}{Final  state}&
                                    &\multicolumn{2}{c}{Initial state}
                                          &  Amplitude    \\
\multicolumn{2}{c}{$\Delta(J,j_{V},\frac{1}{2}),\Delta(j_{V},\ell_V,1)$}
 & &\multicolumn{2}{c}{$\Delta(J,j_{\gamma},\frac{1}{2}),
\Delta(j_{\gamma},\ell_{\gamma},1)$}
  & $M(j_{\gamma}=\ell_{\gamma})$\\
   $\ell_V$                         &   $j_{V}$  & J
 &  $j_{\gamma}$         &   Parity
 & $E(j_{\gamma}=\ell_{\gamma}\pm 1)$\\
\tableline
\\
\noindent{\smallskip}
   $L$            &  $ L+1$  & $j_V+1/2$ &  $J-1/2$  & $(-1)^{j_{\gamma}}$ &
$E_{L+}^{+}$\\
\vspace{.05in}
                      &       &    &  $J+1/2$  & $-(-1)^{j_{\gamma}}$  &
$M_{L+}^{+}$\\
\tableline\\
\noindent{\smallskip}
   $L$            &  $ L  $  & $j_V+1/2$ &  $J-1/2$  & $-(-1)^{j_{\gamma}}$  &
$M_{L0}^{+}$\\
\vspace{.05in}
                      &      &   &  $J+1/2$  & $(-1)^{j_{\gamma}}$  &
$E_{L0}^{+}$\\
\tableline\\
\noindent{\smallskip}
$L$               &  $ L-1$  & $j_V+1/2$ & $J-1/2$  & $(-1)^{j_{\gamma}}$  &
$E_{L-}^{+}$\\
\vspace{.05in}
                      &      &     &  $J+1/2$  & $-(-1)^{j_{\gamma}}$  &
$M_{L-}^{+}$\\
\tableline\\
\noindent{\smallskip}
   $L$            &  $ L+1$  & $j_V-1/2$ & $J-1/2$  & $-(-1)^{j_{\gamma}}$  &
$M_{L+}^{-}$\\
\vspace{.05in}
                      &      &     &  $J+1/2$  & $(-1)^{j_{\gamma}}$  &
$E_{L+}^{-}$\\
\tableline\\
\noindent{\smallskip}
   $L$            &  $ L  $  & $j_V-1/2$ & $J-1/2$  & $(-1)^{j_{\gamma}}$  &
$E_{L0}^{-}$\\
\vspace{.05in}
                      &       &      &  $J+1/2$  & $-(-1)^{j_{\gamma}}$  &
$M_{L0}^{-}$\\
\tableline\\
\noindent{\smallskip}
   $L$            &  $ L-1$  & $j_V-1/2$ & $J-1/2$  & $-(-1)^{j_{\gamma}}$  &
$M_{L-}^{-}$\\
\vspace{.05in}
                      &      &      &  $J+1/2$  & $(-1)^{j_{\gamma}}$  &
$E_{L-}^{-}$
\end{tabular}
\label{table1}
\end{table}
\widetext
\begin{table}
\caption{ Resonance behavior of observables near threshold. Here each column
indicates the angular behavior of observables when
{\em only one amplitude} is nonzero.}
\begin{tabular}{ccccc}
\noindent{\smallskip}
Observable & $C^{\gamma,N,N',V}_{l,k,k',ij}$ & $E_{0+}^{-}$
& $E_{0+}^{+}$ & $M_{0+}^{+}$\\
\tableline\\
\noindent{\smallskip}
Cross-Section & I &   flat   &    flat       &     flat       \\
\tableline\\
\noindent{\smallskip}
Single Spin&$C^{V}_{x'x'}$& 0 & $2-3\,{\rm cos}^2\theta$ &
 $3\,{\rm cos}^2\theta-2$\\
\vspace{.05in}
 $''$ &$C^{V}_{y'y'}$& 0 & flat$<0$ &   flat$>0$     \\
\vspace{.05in}
  $''$&$C^{V}_{z'z'}$& 0 & $3\,{\rm cos}^2\theta-1$ & $1-3\,
{\rm cos}^2\theta$   \\
\vspace{.05in}
 $''$ &$C^{V}_{x'z'}$& 0 & ${\rm sin}(2\,\theta)$ & $-{\rm sin}(2\,\theta)$\\
\tableline\\
\noindent{\smallskip}
Beam Target&$C^{\gamma N}_{z,z}$ & flat$<0$ &  flat$>0$ & flat$<0$     \\
\tableline\\
\noindent{\smallskip}
Target Recoil &$C^{NN'}_{x,x'}$& 0 & ${\rm cos}\,
\theta$ &$-{\rm cos}\,\theta$\\
\vspace{.05in}
$''$  &$C^{NN'}_{x,z'}$& 0 & $-{\rm sin}\,\theta$ &$ {\rm sin}\,\theta$ \\
\vspace{.05in}
 $''$ &$C^{NN'}_{z,x'}$&${\rm sin}\,\theta$ & ${\rm sin}\,\theta$ & 0 \\
\vspace{.05in}
$''$  &$C^{NN'}_{z,z'}$&${\rm cos}\,\theta$ & ${\rm cos}\,\theta$ & 0 \\
\vspace{.05in}
$''$  &$C^{NN'}_{y,y'}$& 0 & flat$>0$ & flat$<0$ \\
\tableline\\
\noindent{\smallskip}
Beam Recoil &$C^{\gamma N'}_{z,x'}$ & -${\rm sin}\,\theta$ & ${\rm sin}\,
\theta$ & ${\rm sin}\,\theta$ \\
\vspace{.05in}
$''$  &$C^{\gamma N'}_{z,z'}$ & -${\rm cos}\,\theta$ & ${\rm cos}\,
\theta$ & ${\rm cos}\,\theta$ \\
\tableline\\
\noindent{\smallskip}
Beam V.Meson&$C^{\gamma V}_{z,x'}$ &${\rm sin}\,\theta$  & ${\rm sin}
\,\theta$ & ${\rm sin}\,\theta$  \\
\vspace{.05in}
$''$  &$C^{\gamma V}_{z,z'}$ &${\rm cos}\,\theta$ & ${\rm cos}\,
\theta$ & ${\rm cos}\,\theta$  \\
\vspace{.05in}
$''$  &$C^{\gamma V}_{x,x'x'}$ & 0 & ${\rm cos}^2\theta$ &
$-{\rm cos}^2\theta$   \\
\vspace{.05in}
$''$  &$C^{\gamma V}_{y,x'y'}$ & 0 & ${\rm cos}\,\theta$ &
$-{\rm cos}\,\theta$   \\
\vspace{.05in}
$''$  &$C^{\gamma V}_{x,x'z'}$ & 0 & $-{\rm sin}(2\,\theta)$
 & ${\rm sin}(2\,\theta)$ \\
\vspace{.05in}
$''$  &$C^{\gamma V}_{x,y'y'}$ & 0 &  flat$<0$ & flat$>0$      \\
\vspace{.05in}
$''$  &$C^{\gamma V}_{y,y'z'}$ & 0 & $-{\rm sin}\,\theta$ &
 ${\rm sin}\,\theta$ \\
\vspace{.05in}
$''$  &$C^{\gamma V}_{x,z'z'}$ & 0 & ${\rm sin}^2\theta$ &
 $-{\rm sin}^2\theta$\\
\tableline\\
\noindent{\smallskip}
Target V.Meson& $C^{NV}_{x,x'}$ & 0 & ${\rm cos}\,\theta$
 &$-{\rm cos}\,\theta$ \\
\vspace{.05in}
$''$  &$C^{NV}_{x,z'}$ & 0 & $-{\rm sin}\,\theta$ & ${\rm sin}\,\theta$  \\
\vspace{.05in}
$''$  &$C^{NV}_{y,y'}$ & 0 & flat$>0$ & flat$<0$\\
\vspace{.05in}
$''$  &$C^{NV}_{z,x'}$ &$-{\rm sin}\,\theta$ & ${\rm sin}\,\theta$ & 0   \\
\vspace{.05in}
$''$  &$C^{NV}_{z,z'}$ &$-{\rm cos}\,\theta$ & 0 & 0 \\
\tableline\\
\noindent{\smallskip}
Recoil V.Meson &$C^{N'V}_{x',x'}$ & flat$<0$ & $4-3\,
{\rm cos}^2\theta$\ & 0  \\
\vspace{.05in}
$''$  &$C^{N'V}_{y',y'}$ & flat$<0$ & flat$>0$ & flat$>0$ \\
\vspace{.05in}
$''$  &$C^{N'V}_{z',z'}$ & flat$<0$ & $3\,{\rm cos}^2\theta+1$\ &
${\rm sin}^2\,\theta$ \\
\vspace{.05in}
$''$  &$C^{N'V}_{z',x'}$ & 0 & ${\rm sin}(2\,\theta)$ &
 ${\rm sin}(2\,\theta)$ \\
\vspace{.05in}
$''$  &$C^{N'V}_{x',z'}$ & 0 & ${\rm sin}(2\,\theta)$ &
$-{\rm sin}(2\,\theta)$
\end{tabular}
\label{table2}
\end{table}

\end{document}